\documentclass[prd,aps,a4paper,twocolumn,eqsecnum,nofootinbib,floatfix]{revtex4}  %

\newif\ifusesec
\usesectrue  
   
\usepackage{graphicx} 
\usepackage{amsmath,amsfonts,amssymb}
\usepackage[section]{placeins}
\usepackage{appendix}
\usepackage{mathtools}
\usepackage[caption=false,justification=justified]{subfig}
\usepackage{xcolor}

\newcommand{\beq}{\begin{equation}}
\newcommand{\eeq}{\end{equation}}
\newcommand{\bea}{\begin{eqnarray}}
\newcommand{\eea}{\end{eqnarray}}
\begin{document}

\title{Radial fall: the gravitational waveform up to the second-and-half Post-Newtonian order}

\author{Donato Bini$^{1}$, Giorgio Di Russo$^{2}$}
  \affiliation{
$^1$Istituto per le Applicazioni del Calcolo ``M. Picone,'' CNR, I-00185 Rome, Italy\\
$^2$School of Fundamental Physics and Mathematical Sciences, Hangzhou Institute for Advanced Study, UCAS, Hangzhou 310024, China
}

\date{\today}

\begin{abstract}
We consider an application of the Multipolar Post Minkowskian formalism to the case of a two-body system in radial fall. We compute, within the post-Newtonian approximation, the associated gravitational waveform  reaching the 2.5 Post-Newtonian accuracy level. At this level the presence of a radiation-reaction force manifests,  modifying the fall with a corresponding bremsstrahlung radiation. We evaluate then all emissions: energy, angular momentum (vanishing identically) and linear momentum. We also evaluate the (nonlocal)   inertial forces contributions appearing (at the next PN order, 4.5PN) in the center-of-mass due to the losses paving the way for future more accurate computations.
\end{abstract}

\maketitle

\section{Introduction}
\label{Intro}

The study of matter undergoing radial infall into a black hole, and the associated emission of gravitational radiation, has a long history. Pioneering investigations of this problem were carried out in the 1970s , when Zerilli \cite{Zerilli:1970wzz} and  Davis, Ruffini, Press, Price and Tiomno \cite{Davis:1971gg,Davis:1972ud,Ruffini:1973ky}  first computed the radiated energy, both numerically and analytically, at leading order in the high-frequency regime for the even-parity sector of Schwarzschild metric perturbations, described by the Zerilli equation \cite{Zerilli:1970se}.

Other numerical studies of massive particle motion in Schwarzschild spacetime, in particular within the Teukolsky formalism \cite{Teukolsky:1973ha,Press:1973zz,Teukolsky:1974yv}, were carried out by Detweiler for various values of the angular momentum, including the case of vanishing angular momentum \cite{Detweiler:1979xr}. Furthermore, the radial fall in the Kerr geometry was studied numerically in \cite{Sasaki:1981sx}. 

From the two-body system perspective, radial collisions (often referred to as \lq \lq head-on" collision processes, or \lq\lq direct radial plunge") were studied in \cite{Anninos:1993zj,Simone:1995qu}, where numerical results were compared with the leading-order multipolar expansion.

Pedagogical discussions of this problem can be found in \cite{Maggiore:2007ulw,Maggiore:2018sht}. In all these works, the leading Newtonian order contribution to the emission is derived under the following assumptions. First, gravitational waves (GWs) production is computed in linearized theory, neglecting backreaction on the Schwarzschild background. Second, Newtonian equations of motion are used instead of the relativistic ones. Third, it is assumed that most of the radiation is emitted while the particle is non relativistic, so that the quadrupole formula remains valid.

More recent and refined numerical studies of the radial infall of massive particles in Schwarzschild and Kerr spacetimes have been presented in \cite{Oliveira:2018pnx,Mitsou:2010jv,Barausse:2021xes,DeAmicis:2024eoy,Scheel:2025jct}. These works investigated not only the inspiral phase but also the ringdown, in particular the emission of characteristic quasi-normal mode (QNM) frequencies. The study of QNMs, together with tidal deformations and wave amplification factors, has received considerable attention recently, motivated by the GW event GW250114 \cite{LIGOScientific:2025rid}, which was detected with a particularly high signal-to-noise ratio. The theoretical and numerical techniques to compute these quantities are now well-known, ranging from the pioneering works of Chandrasekhar \cite{Chandrasekhar:1975zza} to some recent developments based on conformal field theory (CFT) methods \cite{Berti:2009kk,Aminov:2020yma}, and have been applied even beyond the context of astrophysical black holes \cite{Bianchi:2021xpr,Bianchi:2021mft,Bianchi:2022qph,Bianchi:2023sfs,DiRusso:2024hmd,Bena:2024hoh,Cipriani:2024ygw,Fioravanti:2021dce,DiRusso:2025qpf}.

Finally, the first analytical study of this process within the framework of gravitational self-force theory \cite{Poisson:2003nc,Barack:2018yvs}, for both scalar and curvature perturbations, was performed very recently in \cite{Bini:2026cpw}. In that work, the solutions of the homogeneous Teukolsky equations were computed using the Mano-Suzuki-Takasugi method \cite{Mano:1996vt}. This analytical framework has recently been extended to solitonic solutions of five-dimensional Einstein–Maxwell theory that behave as Schwarzschild mimickers, known as Topological Stars \cite{Bah:2020pdz,Bianchi:2024vmi,Bianchi:2024rod,DiRusso:2025lip,Bianchi:2025aei,Bini:2025ltr,Bianchi:2025uis,Heidmann:2025pbb}, as well as to W solitons \cite{Dima:2025tjz,Bianchi:2025ydq}. However, other analytical approaches based on CFT techniques have been developed and tested against established numerical and semi-analytical results \cite{Fucito:2024wlg,Cipriani:2025ikx,Cipriani:2026myb,Cipriani:2026xmx}. The equivalence between the MST approach and the CFT-inspired techniques has been discussed in \cite{Bini:2025bll}, where, following \cite{Ivanov:2025ozg}, the equivalence between the renormalized angular momentum and the so-called Seiberg-Witten fundamental cycle was explicitly demonstrated. 

Although the radial infall emission process has been extensively studied, the vast majority of approaches to this problem rely on numerical or semi-analytical techniques, typically at lowest order in some approximation scheme. This is due to the peculiar nature of the orbit along which the massive particle moves. For simplicity, one can consider the particle as starting its motion from rest at infinity. During the initial phase of the motion, the particle moves in a weak gravitational field, where the Post-Newtonian (PN) approximation (i.e., non-relativistic motion in a weak field) is valid.

After a certain amount of time, however, the particle necessarily enters the strong field region, and here all the tools available in perturbation theory cannot be used more implying that the description can only be continued within numerical relativity. In other words, the situation of a radial fall is quite different from both a coalescing process (where one imagine that the particle may spiral for many cycles around the black hole before meeting a strong field region) and from a scattering process (which also might occur far away the black hole location where PN approximation works well, see, e.g. Refs. \cite{Bini:2012ji,Bini:2017wfr,Bini:2021gat} for genuine PN computations). 

As already anticipated, a possibility to perform a (partial) analytical study is then to stop the PN description of the radial fall as soon as the particle approaches the black hole horizon. This is equivalent to introduce a temporal scale (say $t_{\rm max}$) beyond which PN formalism cannot be used more. 

With this limitation in mind, i.e., within the time interval ($t>t_{\rm max}$ with our parametrization of the orbit, see below), one can perform several analytic interesting computations: the orbit (including radiation-reaction effects), the waveform (radiation-reacted too), the various losses of energy and momenta. Our aim here is to collect all these pieces of information in PN sense and prepare the land for future checks against a fully numerical study. However, as we shall see below in detail, numerical studies show that the energy radiated to infinity is emitted is peaked at $M\omega \sim 0.36$ (i.e., when the particle is still in the weak-field regime). This behavior is captured by the PN expansion.

The paper is organized as follows. In Sec. II we recall all the building blocks necessary to compute the waveform at the 2.5PN accuracy level.
In Sec. III we discuss the relative motion in the presence of a radiation-reaction acceleration at the 2.5PN accuracy while in Sec. IV 
we explicitly evaluate all the needed multipoles. Finally in Sec. V we discuss the appearance of inertial forces in the CM frame (at 4.5PN) which will be necessary to be taken into account as soon as the 4.5PN accuracy will be reached (which, however, is not the case with the present paper).

Our metric signature convention here is the mostly-positive one, $-+++$. We often use units such that $c=1=G$ and denote $\eta=1/c$ as a place-holder for PN expansions.
Different choices are eventually specified in the text.
 
\section{Waveform within the MPM formalism and at the 2.5PN accuracy level}

The  transverse-traceless (TT) asymptotic waveform  $\lim_{R\to \infty}( R \, h_{ij}^{\rm TT})$ is conveniently
summarized in the complex (metric) quantity
\bea
h_c(t_r,\theta,\phi) &=&\lim_{R\to \infty}(R( h_+ -  i h_\times))\nonumber\\
&=&\lim_{R\to \infty} \bar m^{\mu } \bar m^{\nu }R\, h_{\mu \nu}\,.
\eea
Here, 
\beq \label{tret}
t_r = t-\frac{r}{c}-\frac{2 G{\cal M} }{c^3}\ln \frac{r}{r_0},
\eeq
is the  Bondi-type  retarded time (depending on the
arbitrary length scale $r_0$), and $\bar m^{\mu }$ a null polarization vector chosen so 
to have only CM spatial components $\bar m^j$, see Refs. \cite{Bini:2023fiz,Bini:2024rsy,Bini:2024ijq}, orthogonal to the observation direction denoted by $n^i$.
More precisely, our choice for the observation direction and the polarization vector is the following
\bea
n^{i}&=& (s_\theta c_\phi, s_\theta s_\phi, c_\theta)\,,\nonumber\\
e_{\hat \theta}^i&=&\frac{\partial n^{i}}{\partial \theta}\,,\quad
e_{\hat \phi}^i=\frac{1}{s_\theta}\frac{\partial n^{i}}{\partial \phi}\,,\quad\bar{m}^i=\frac{1}{\sqrt{2}}(e_{\hat \theta}^i-i e_{\hat \phi}^i)\,,\nonumber\\
\eea
(where $[\cos \alpha, \sin\alpha]=[c_\alpha,s_\alpha]$), with respect to a Cartesian-like coordinate system with the origin in the center-of-mass (CM) of the system.
Explicitly, the components of $\bar{m}^i$ read
\beq
\bar{m}^i {=} \frac{1}{\sqrt{2}}\left((c_\theta c_\phi{+}i s_\phi),(c_\theta s_\phi{-}i c_\phi),{-}s_\theta\right)\,.
\eeq

The Multipolar Post Minkowskian  (MPM) formalism computes  the  multipolar decomposition  of $h_c(t_r,\theta,\phi)$.
More precisely, introducing the rescaled complex time-domain waveform $W$ such that 
 \beq
 h_c(t_r,\theta,\phi) \equiv 4 G \eta^4  W(t_r,\theta,\phi)\,,
 \eeq
one has the following multipolar decomposition for $W$
\bea
\label{W_deco}
W(t_r,\theta,\phi)&=& U_2+ \eta (V_2 +U_3) + \eta^2 (V_3+U_4)\nonumber\\ 
&+& \eta^3 (V_4+U_5)+ \cdots \,.
\eea
Here, each $U_\ell$ (respectively, $V_\ell$) denotes an even-parity (respectively,  odd-parity) $2^\ell$ radiative multipole contributions.
In the MPM formalism they are expressed in terms 
of symmetric-trace-free (STF) Cartesian-type tensors of order $\ell$ (the radiative multipole moments  $U_{ i_1 i_2 \cdots i_{\ell}}(t_r)$
and  $V_{ i_1 i_2 \cdots i_{\ell}}(t_r)$) according to the following definitions
\bea\label{UV}
U_\ell(t_r,\theta,\phi) &=& \frac{1}{\ell!} \bar m^{i} \bar m^{j } n^{i_1} n^{i_2} \cdots n^{i_{\ell-2}} U_{i j i_1 i_2 \cdots i_{\ell-2}}(t_r)\,, \nonumber\\
V_\ell(t_r,\theta,\phi) &=& - \frac{1}{\ell!}\frac{2\ell}{\ell+1} \bar m^{i} \bar m^{j } n^c  n^{i_1} n^{i_2} \cdots n^{i_{\ell-2}}\times \nonumber\\
&&  \epsilon_{cd i}V_{j d i_1 i_2 \cdots i_{\ell-2}}(t_r)\,.
\eea
The PN-matched MPM formalism computes the  radiative moments (observed at future null infinity) as nonlinear retarded functionals
(obtained by iteratively solving  Einstein's vacuum equations in the exterior zone) of two other types of multipole moments: (i) two sequences of \lq\lq source moments," 
\beq
 I_{L}(t_r)\,,\qquad J_{L}(t_r)\,,
\eeq 
with $L=i_1 i_2 \cdots i_{\ell}$ a multi-index notation,
together with four sequences of \lq\lq gauge moments":  
\bea
&& W_{L}(t_r)\,,\quad X_{L}(t_r)\,,\quad Y_{L}(t_r)\,,\quad  Z_{L}(t_r)\,.
\eea
The source and gauge moments are then computed  as explicit expressions of the source degrees of freedom: namely, this means explicit expressions in terms of the positions and velocities. We shall compute all the needed multipole moments in the incoming (i.e. at the starting of the radial fall) CM frame of the system, and in modified harmonic coordinates \cite{Blanchet:2013haa}.

The frequency-domain waveform $W(\omega,  \theta,\phi) $ is then obtained by Fourier-transforming (over the retarded time variable) 
the radiative moments,
\bea
U_{i_1 i_2 \cdots i_{\ell}}(\omega)= \int_{- \infty}^{+ \infty} dt_r e^{i \omega t_r} U_{i_1 i_2 \cdots i_{\ell}}(t_r)\,, \nonumber \\
V_{i_1 i_2 \cdots i_{\ell}}(\omega)= \int_{- \infty}^{+ \infty} dt_r e^{i \omega t_r} V_{i_1 i_2 \cdots i_{\ell}}(t_r)\,.
\eea
This  leads to the following multipole expansion of the frequency-domain complex waveform 
$h_c(\omega,\theta,\phi) \equiv 4 G \eta^4 W(\omega,  \theta,\phi)$:
\bea
\label{Wom_th_phi}
W(\omega,  \theta,\phi) &\equiv& U_2(\omega,\theta,\phi)\nonumber\\
&+& \eta (V_2(\omega,\theta,\phi) +U_3(\omega,\theta,\phi))\nonumber\\ 
&+& \eta^2 (V_3(\omega,\theta,\phi)+U_4(\omega,\theta,\phi))\nonumber\\ 
&+& \eta^3 (V_4(\omega,\theta,\phi)+U_5(\omega,\theta,\phi))\nonumber\\
&+& \eta^4 (V_5(\omega,\theta,\phi)+U_6(\omega,\theta,\phi))\nonumber\\
&+& \eta^5 (V_6(\omega,\theta,\phi)+U_7(\omega,\theta,\phi))\nonumber\\
&+& O(\eta^6)\,,
\eea
where, for instance, the quadrupole contribution reads
\beq
U_2(\omega,\theta,\phi)=  \frac{1}{2!} \bar m^{i} \bar m^{j } U_{i j}(\omega)\,.
\eeq
To compute $U_2$ (and then $W$)  at the 2.5PN accuracy level we recall that
\bea
\label{U_def}
U_{ij} &=& M_{ij}^{(2)} + \eta^3 U_{ij}^{\rm 1.5PN}  +\eta^5 U_{ij}^{\rm 2.5PN}\,,
\eea
with
\bea
\label{blocks_0}
M_{ij} &=& 
I_{ij}
	+ 4G \eta^5 (W^{(2)} I^{}_{ij}- W^{(1)} I^{(1)}_{ij})\,.
\eea
The other terms in Eq. \eqref{U_def} at our accuracy  read 
\bea
\label{lista_up_2.5pn}
U_{ij}^{\rm 1.5PN}(t_r) &=&  2 G {\cal M}\eta^3  \int_0^{+\infty} d\tau\,  M_{ij}^{(4)}(t_r-\tau)\ln \left(\frac{\tau}{C_{I_2}}\right)\,,\nonumber\\
\eea
with with ${\cal M} \equiv \frac{E}{c^2}$ denoting the total ADM mass in the CM system and
\beq
C_{I_2}=\frac{2 b_0 e^{-11/12}}{c}\,.
\eeq
Here $U_{ij}^{\rm 1.5PN}(t_r)$ has to be computed at 1PN fractional accuracy, while the
additional contributions (genuine 2.5PN contributions, see Eq. \eqref{U_def}) are given by
\bea
U_{ij}^{{\rm 2.5PN}}(t_r) &=&  G\eta^5  \left( 
U_{ij}^{{\rm 2.5PN}\rm (mem)}
+U_{ij}^{{\rm 2.5PN} I_2I_2}\right.\nonumber\\
&+&\left. \text{U}_{ij}^{ {\rm 2.5PN}I_2J_1}\right)\,,
\eea
where
\bea
U_{ij}^{{\rm 2.5PN} \rm  (mem)}(t_r)&=&- \frac{2}{7} \int_0^{+\infty} \! d\tau\!\left[ M^{(3)}_{a\langle i} M^{(3)}_{j\rangle a}\right]\!(t_r-\tau)\,,\nonumber\\
U_{ij}^{{\rm 2.5PN}\, I_2I_2}(t_r)&=&  \frac{1}{7}\, M^{(5)}_{a\langle i} M^{}_{j\rangle a} - \frac{5}{7} \, M^{(4)}_{a\langle i} M^{(1)}_{j\rangle a}\nonumber\\
&-& \frac{2}{7}\, M^{(3)}_{a\langle i} M^{(2)}_{j\rangle a}  
\,,\nonumber\\
U_{ij}^{{\rm 2.5PN}I_2J_1}(t_r)&=&  \frac{1}{3}\epsilon^{}_{ab\langle i} M^{(4)}_{j\rangle a}   S^{}_{b}  \,.
\eea
Finally, $U_{ij}^{{\rm 2.5PN} \rm  (mem)}$, $U_{ij}^{{\rm 2.5PN}\, I_2I_2}$ and $U_{ij}^{{\rm 2.5PN}I_2J_1}$ have to be computed at the Newtonian accuracy level.
To proceed further let us note that
1) due to the property \eqref{blocks_0} in all terms in Eq. \eqref{lista_up_2.5pn} one can replace $M_{ij}$ as $I_{ij}$;
2) in the radial motion case we are going to discuss  $J_a=0=S_a$ (implying $U_{ij}^{{\rm 2.5PN}I_2J_1}=0$) and $I_{ij}=f(t)q_{ij}$ with $q_{ij}={\rm diag}\left[1,-\frac12, -\frac12\right]$ a constant matrix, which will imply special simplifications, as for example,
\beq
q_{\langle ij\rangle }^2= q_{ij}^2-\frac12 \delta_{ij}=\frac{1}{2}q_{ij}\,,
\eeq
and ${\rm Tr}(q_{ij}^2)=\frac32$, besides the obvious advantage of simplifying most of the tensorial relations, e.g., 
\beq
M^{(5)}_{a\langle i} M^{}_{j\rangle a}{=} f^{(5)}(t)f(t) q_{\langle ij\rangle }^2{=}\frac12 f^{(5)}(t)f(t)q_{ij} \,. 
\eeq

\section{Motion in the presence of a radiation-reaction acceleration at the 2.5PN accuracy}

The relative motion of the system with respect of the CM is described by the relative acceleration ${\mathbf a}={\mathbf a}_1-{\mathbf a}_2$ (with ${\mathbf a}_a$, $a=1,2$ the accelerations of the two bodies). Its expression is given by
\bea
{\mathbf a}={\mathbf a}_{\rm N}+{\mathbf a}_{\rm 1PN}+ {\mathbf a}_{\rm 2PN}+{\mathbf a}_{\rm 2.5PN}\,,
\eea 
where  ${\mathbf a}_{\rm N}$, ${\mathbf a}_{\rm 1PN}$, ${\mathbf a}_{\rm 2PN}$ can be found in 
Eqs. (358a), (358b) and (358c) of Ref. \cite{Blanchet:2013haa}. The 2.5PN acceleration in harmonic coordinates originally due to Damour and Deruelle \cite{Damour:1981bh}
(explicitly displayed below for convenience with $G=1$) reads
\beq
{\bf a}_{\rm 2.5PN}=  -\frac{8}{5}\nu\frac{M^2}{r^3}(-A_{2.5}\dot r {\bf n}+B_{2.5}{\bf v}) \,,
\eeq
with
\bea
A_{2.5}&=&3v^2+\frac{17M}{3r }\,,\nonumber\\ 
B_{2.5}&=& v^2+3\frac{M}{r}\,.
\eea
Solving for the motion corresponding to \lq\lq release from rest at infinity" in Cartesian coordinates we find the following solution
\bea
\label{orbit}
x_p(t)&=&\frac{3^{2/3}
    M^{1/3} t^{2/3}}{2^{1/3}}+\frac{5}{2} \eta ^2 (\nu -2)  M\nonumber\\
   &+&\eta ^4 \frac{\left(5 \nu^2-19\nu +48\right) 
   M^{5/3}}{2\cdot 6^{2/3}t^{2/3}}\nonumber\\
   &+&\eta ^5\epsilon \frac{64  \nu   M^2
   }{63 t}+O(\eta^6)\,,\nonumber\\
y_p(t)&=& O(\eta^6)\,,
\eea
where $\eta=\frac{1}{c}$ is a PN place-holder and $\epsilon$ is a radiation-reaction place-holder and the masses $m_1$ and $m_2$ ($m_1>m_2$) of the bodies form the standard combinations: $m_1+m_2=M$ (total mass), $\mu=m_1m_2/(m_1+m_2)$ (reduced mass), $q=m_2/m_1$ (mass ratio), $\nu=m_1m_2/(m_1+m_2)^2$ (symmetric mass ratio). The radial fall is assumed to be along the $x-$axis.

\begin{figure}
\includegraphics[scale=0.8]{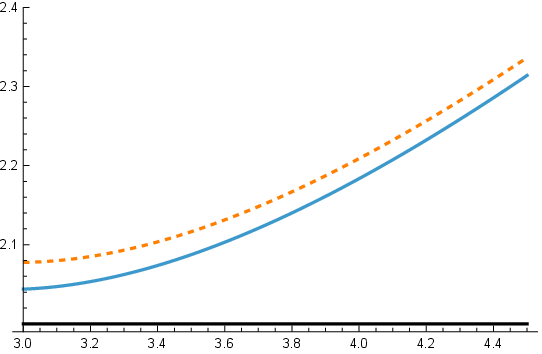}
\caption{\label{fig:1} We show the comparison between the conservative trajectory ($\epsilon=0$ continuous blue line) with the trajectory corrected with the radiation reaction effect ($\epsilon=1$ dashed orange line) having chosen $\nu=0.1$. [Note that the expression of the radiation-reaction force (magnitude of the radial component) along the motion reduces to $|{\bf a}_{\rm 2.5PN}^r|=\frac{128 M^2\nu}{81 t^3}$]. Superposed to the other curves we show (black online) the position of the horizon ($r=2$ in units of $M$) when $\nu=0$, as a reference line.}
\end{figure}

The behavior of the conservative trajectory compared with the trajectory corrected with the radiation reaction is displayed in Fig.\ref{fig:1}.

The conservative part of the orbit was found in Ref. \cite{Bini:2026cpw}. Note that in Ref. \cite{Bini:2026cpw} a different parametrization was used. In the present case, the particle starts its motion at infinity at $t=\infty$, at rest, and then, as time decreases, it moves towards the horizon. The final approach cannot   be followed within a PN analysis, since the PN expansion breaks down when the particle enters the strong-field regime. This argument, as well as the limitations of this kind of computation, are extensively discussed in \cite{Maggiore:2007ulw}. The radiation-reaction part has been obtained by imposing that it vanishes at $t\to \infty$ (where the motion starts), as it is customary. 
Often we will introduce a dimensionless time parameter $T=t/M$.

\section{Expressing the radiative moments in terms of the source multipole moments}
Let us assume that the two bodies motion is confined to the $\langle x,y\rangle$ plane
\beq
x^i=(x (t),y (t),0),\quad v^i=(\dot x(t),\dot y(t),0).
\eeq
The quadrupole mass multipole is then given by
\bea
I_{ij}&=&\nu M \Bigg\{\Bigg[A{-}\frac{24}{7}\nu\eta^5\frac{G^2M^2}{r^2}\dot r\Bigg] x_{\langle i}x_{j\rangle}{+}B\eta^2\,r^2 v_{\langle i}v_{j\rangle}\nonumber\\
&{+}&2\Bigg[C\eta^2\,r \dot r{+}\frac{24}{7}\nu \eta^5\frac{G^2M^2}{r}\Bigg] x_{\langle i}v_{j\rangle}\Bigg]\Bigg\}{+}O\left(\eta^6\right)\,,\nonumber\\
\eea
where $A$, $B$, $C$ can be found in Eqs. (3.2a,b,c)  of Ref. \cite{Arun:2007sg}.

In the present case, inserting the orbit \eqref{orbit} in the expression of $I_{ij}$ one finds
\bea
I_{ij}=f_2(t)q_{ij}\,,\qquad q_{ij}={\rm diag}\left[1,-\frac12, -\frac12\right]\,,
\eea
with
\bea\label{f2}
\frac{f_2(t)}{\nu M^3}&=&6^{1/3}  T^{4/3}\nonumber\\
&+&\eta^2 \frac{2^{2/3}T^{2/3}}{3^{1/3}}\left(-\frac{66}{7}+\frac{16\nu}{7}\right)\nonumber\\
&+&\eta^4\left(\frac{16 \nu ^2}{21}-\frac{28
   \nu }{3}+\frac{1327}{42}\right)\nonumber\\
   &+&\frac{272\cdot 2^{2/3}\eta^5\nu}{63\cdot 3^{1/3}T^{1/3}}+O\left(\eta^6\right)\,,
\eea 
where we introduced the dimensionless variable $T=\frac{t}{M}$ and took into account the big simplifications due to the diagonal character of $I_{ij}$.
Explicitly, the (mass-type) radiative quadrupole moments are then given by
\bea
U_{ij}(t)&=&I^{(2)}_{ij}(t)\nonumber\\
&+&2M\eta^3\int_0^\infty d\tau \ln \left(\frac{\tau}{C_{I_2}}\right)I^{(4)}_{ij}(t-\tau)\nonumber\\
&+&\eta^5\Bigg\{-\frac{2}{7}\int_0^\infty d\tau I^{(3)}_{a\langle i}(t-\tau)I^{(3)}_{j\rangle a}(t-\tau)\nonumber\\
&+&\frac{1}{7}I^{(5)}_{a\langle i}I_{j\rangle a}-\frac{5}{7}I^{(4)}_{a\langle i}I^{(1)}_{j\rangle a}-\frac{2}{7}I^{(3)}_{a\langle i}I^{(2)}_{j\rangle a}\nonumber\\
&{+}&\frac{1}{3}\epsilon_{ab\langle i}I^{(4)}_{j\rangle a}J_b{+}4\Bigg[W^{(2)}I_{ij}{-}W^{(1)}I_{ij}^{(1)}\Bigg]^{(2)}\Bigg\}\nonumber\\
&+&O\left(\eta^6\right)\,,\nonumber\\
\eea
\bea
U_{ijk}(t)&=&I^{(3)}_{ijk}(t)\nonumber\\
&{+}&2M\eta^3\int_0^\infty d\tau I^{(5)}_{ijk}(t{-}\tau)\Bigg[\log\left(\frac{c\tau}{2b_0}\right){+}\frac{97}{60}\Bigg]\nonumber\\
&+&O\left(\eta^5\right)\,,\nonumber\\
U_{ijkl}(t)&=&I^{(4)}_{ijkl}(t)\nonumber\\
&{+}&\eta^3\Bigg\{2M\int_0^\infty d\tau I_{ijkl}^{(6)}(t{-}\tau)\Bigg[\log\left(\frac{c\tau}{2b_0}\right){+}\frac{59}{30}\Bigg]\nonumber\\
&+&\frac{2}{5}\int_0^\infty d\tau\left(I^{(3)}_{\langle ij}I^{(3)}_{kl\rangle}\right)(t-\tau)-\frac{21}{5}I_{\langle ij}I_{kl\rangle}^{(5)}\nonumber\\
&-&\frac{63}{5}I_{\langle ij}^{(1)}I_{kl\rangle}^{(4)}-\frac{102}{5}I_{\langle ij}^{(2)}I_{kl\rangle}^{(3)}\Bigg\}+O\left(\eta^4\right)\,,\nonumber\\
U_{ijklm}&=&I_{ijklm}^{(5)}+O\left(\eta^3\right)\,,\nonumber\\
U_{ijklmn}&=&I_{ijklmn}^{(6)}+O\left(\eta^2\right)\,,\nonumber\\
U_{ijklmnp}&=&I_{ijklmnp}^{(7)}+O\left(\eta\right)\,.\nonumber\\
\eea
From the definition Eq. \eqref{UV} one obtains 
\bea
U_2&=&\nu M (c_\theta
    c_\phi+i s_\phi
   )^2\nonumber\\
   &\times&\Bigg[\frac{1}{6^{2/3} T^{2/3}}+\frac{\eta^2}{6^{1/3}T^{4/3}}\left(\frac{11}{7}-\frac{8 \nu }{21}\right)\nonumber\\
   &+&\frac{\eta^3}{6^{2/3}T^{5/3}}\left(\frac{4 \mathcal{L}}{3}+\frac{2
   \pi }{3 \sqrt{3}}+\frac{1}{9}\right)\nonumber\\
   &+&\frac{2^{2/3}\eta^5}{3^{1/3}T^{7/3}}\Big(\frac{16 \pi  \nu }{63
   \sqrt{3}}-\frac{1658 \nu
   }{1323}-\frac{22 \pi }{21
   \sqrt{3}}+\frac{44}{9}\nonumber\\
   &+&\mathcal{L}\Big(\frac{44}{21}-\frac{32 \nu }{63}\Big)\Big)+O\left(\eta^6\right)\Big]\,,\nonumber\\
   \eea
   \bea
U_3&=&\nu M \Delta c_\phi s_\theta (c_\theta c_\phi+i \,s_\phi)^2\nonumber\\
&\times&\Bigg[\frac{\eta^2}{6^{2/3}T^{5/3}}\left(\frac{20 \nu }{27}-\frac{85}{27}\right)\nonumber\\
&+&\frac{\eta^4}{6^{1/3}T^{7/3}}\left(-\frac{20 \nu ^2}{81}+\frac{214
   \nu }{81}-\frac{211}{27}\right)\nonumber\\
   &+&O\left(\eta^5\right)\Bigg]\,,\nonumber\\
U_4&{=}&\nu M(7c_{2\theta}{-}3{-}14c_{2\phi}s_\theta^2)(c_\theta c_\phi{+}i\, s_\phi)^2\nonumber\\
&\times&\Bigg[\frac{1}{6^{1/3}T^{4/3}}\left(\frac{5}{504}-\frac{5 \nu }{168}\right)\nonumber\\
&{+}&\frac{\eta^3}{6^{1/3}T^{7/3}}\Big(\left(\frac{5}{189}-\frac{5 \nu
   }{63}\right) \mathcal{L}\nonumber\\
   &+&\frac{5 \pi  \nu }{126
   \sqrt{3}}-\frac{89 \nu
   }{882}-\frac{5 \pi }{378
   \sqrt{3}}+\frac{11}{324}\Big)\Bigg]\nonumber\\
   &+&O\left(\eta^4\right)\,,\nonumber\\
   \eea
   \bea
U_5&{=}&\nu M \Delta(2\nu{-}1)(1{+}3c_{2\theta}{-}6c_{2\phi}s_\theta^2)(c_\theta s_\phi{+}i s_\phi)^2\nonumber\\
&\times&c_\phi s_\theta\Bigg[\frac{7}{216\cdot 6^{2/3}T^{5/3}}{-}\frac{\eta^2}{6^{1/3}T^{7/3}}\left(\frac{25 \nu
   }{324}{-}\frac{25}{162}\right)\Bigg]\nonumber\\
   &+&O\left(\eta^3\right)\,,\nonumber\\
U_6&=&O\left(\eta\right)\,,\nonumber\\
U_7&=&\nu M \Delta(\nu-1)(3\nu-1)c_\phi s_\theta(c_\theta c_\phi+i\, s_\phi)^2\nonumber\\
&\times&\frac{1}{67392\cdot 6^{1/3}T^{7/3}}\Bigg[487+44c_{2\theta}+429 c_{4\theta}\nonumber\\
&-&176(7+13c_{2\theta})c_{2\phi}s_\theta^2+1144c_{4\phi}s_\theta^4\Bigg]\nonumber\\
&+&O\left(\eta\right)\,,
\eea
with
\bea
\mathcal{L}&=&\log\left(\frac{2b_0}{3\sqrt{3}ct}\right)\,,\nonumber\\
\Delta&=&\sqrt{1-4\nu}=\frac{m_1-m_2}{M}\,.
\eea
The decomposition of the various multipolar contributions to the waveform is shown in Table \ref{decY}. The relation between the waveform and the energy loss is
\beq
\frac{dE}{dt}=\frac{1}{16\pi}\int d\Omega |\dot h|^2,\qquad \dot h=4\dot {\mathcal W}\,.
\eeq
Therefore, the contribution of the quadrupole to the energy loss is
\begin{widetext}
\bea
\frac{1}{\nu^2}\left.\frac{dE}{dt}\right|_{2}&=&\frac{32\cdot 2^{2/3}}{405
   \cdot3^{1/3} T^{10/3}}+\frac{\eta^2}{T^4}\left(\frac{2816}{2835}-\frac{2048 \nu
   }{8505}\right)+\frac{2^{2/3}\eta^3}{3^{1/3}T^{13/3}}\left(\frac{1312}{3645}+\frac{64 \pi
   }{243 \sqrt{3}}+\frac{128 \mathcal{L}}{243}\right)\nonumber\\
   &{+}&\frac{2^{1/3}\eta^4}{3^{2/3}T^{14/3}}\left(\frac{16384 \nu
   ^2}{59535}{-}\frac{45056 \nu
   }{19845}{+}\frac{30976}{6615}\right){+}\frac{\eta^5}{T^5}\Big[\frac{4096 \pi  \nu }{25515
   \sqrt{3}}{-}\frac{97024 \nu
   }{25515}{-}\frac{5632 \pi
   }{8505
   \sqrt{3}}{+}\frac{18304}{1215}\nonumber\\
   &+&\Big(\frac{22528}{2835}-\frac{16384
   \nu }{8505}\Big)\mathcal{L}\Big]+O\left(\eta^6\right)\,.
   \eea

The final result in time domain reads
\bea
\label{DEDT}
\frac{1}{\nu^2}\frac{dE}{dt}&=&\frac{32\cdot 2^{2/3}}{405
   \cdot 3^{1/3} T^{10/3}}{+}\frac{\eta^2}{T^4}\left(\frac{2816}{2835 }{-}\frac{2048 \nu }{8505 }\right) {+}\frac{\eta^3}{T^{13/3}}\left(\frac{1312\cdot 2^{2/3}}{3645
   \cdot3^{1/3}}{+}\frac{64\cdot 2^{2/3} \pi }{243\cdot
   3^{5/6}}+\frac{128\cdot 2^{2/3}
   \mathcal{L}}{243\cdot3^{1/3}}\right) \nonumber\\
   &+&\frac{\eta^4}{T^{14/3}}\left(\frac{1076224
   \cdot2^{1/3}}{229635\ 3^{2/3}}-\frac{25088\cdot 2^{1/3} \nu
   }{10935\ 3^{2/3}}+\frac{7936\cdot 2^{1/3} \nu
   ^2}{25515\ 3^{2/3}}\right)+\frac{\eta^5}{T^5}\Big(\frac{18304}{1215}-\frac{5632 \pi }{8505 \sqrt{3}}-\frac{97024 \nu }{25515}\nonumber\\
   &+&\frac{4096 \pi  \nu }{25515
   \sqrt{3}}+\Big(\frac{22528}{2835}-\frac{16384 \nu }{8505}\Big)\mathcal{L}\Big)+O\left(\eta^6\right)\,.
\eea
In the limit $\nu=0$ the right-hand-side of Eq. \eqref{DEDT} coincides with the expansion appearing in Eq. 5.36 of \cite{Bini:2026cpw}.
The total emitted energy follows by integrating the PN-expanded result, Eq. \eqref{DEDT}, over the PN-validity time interval $t\in (t_{\rm max}, +\infty)$ (with our choice of parametrization of the orbit), as explained above.
Similarly, in the frequency domain we have
\bea
\frac{1}{\frac{2}{3}\pi^2\nu^2M^2}\frac{dE}{d\omega}&{=}&\frac{12\cdot 6^{2/3}  \Gamma
   \left(\frac{4}{3}\right)^2}{5\pi }(M \omega)^{4/3}{+}\eta^2\Big(\frac{176\sqrt{3}}{35}{-}\frac{128\nu}{35\sqrt{3}}\Big) (M\omega)^2{-}\frac{24}{5}\eta^3 6^{2/3}  \Gamma
   \left(\frac{4}{3}\right)^2(M \omega)^{7/3}\nonumber\\
   &{+}&\eta^4\frac{2^{1/3} \Gamma\left(\frac{2}{3}\right)^2}{3^{2/3} \pi }\Big(\frac{248 \nu^2}{35}{-}\frac{784 \nu }{15}{+}\frac{33632}{315}\Big)(M\omega)^{8/3}  {+}\eta^5\Big(\frac{256 \pi  \nu }{35\sqrt{3}}{+}\frac{48 \nu}{49}{-}\frac{352 \sqrt{3} \pi}{35}\Big)(M\omega)^3\nonumber\\
   &{+}&O\left(\eta^6\right)\,,
\eea
\end{widetext}
where the relevant integrals appearing in the Fourier transforms computations are given by
\bea
\mathcal{J}_s(\omega)&=&\int_{-\infty}^\infty dt e^{i\omega t}t^s\nonumber\\
&=&2e^{\frac{i\pi s}{2}}|\omega|^{-1-s}\Gamma(1+s)(\Theta(\omega)-1)\sin(\pi s)\,,\nonumber\\
\mathcal{J}^{\rm log}_s(\omega)&=&\int_{-\infty}^\infty dt e^{i\omega t}t^s\log(t)=\frac{d\mathcal{J}_s(\omega)}{ds}\nonumber\\
&=&e^{\frac{i \pi  s}{2}} (\theta
   (\omega )-1) \Gamma (s+1) |
   \omega | ^{-s-1} \nonumber\\
   &\times&\left(i \sin
   (\pi  s) \left(\pi +i \log
   \left(\omega
   ^2\right)\right)+2 \pi  \cos
   (\pi  s)\right.\nonumber\\
   &+&\left.2 \sin (\pi  s) \psi
   ^{(0)}(s+1)\right)\,.
\eea
\begin{figure}
\includegraphics[scale=0.8]{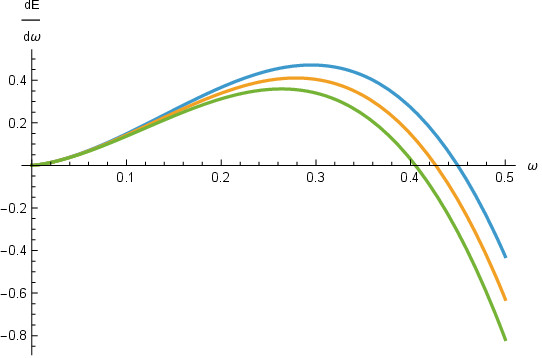}
\caption{\label{fig:2}  The behaviour of $\frac{dE}{d\omega}$ as a function of $\omega$, for $M=1$ and $\nu= [0\, \text{(blue online)},\frac18\, \text{(yellow online)},\frac14\, \text{(green online)}]$.
}
\end{figure}

The loss of angular momentum, defined as
\bea
\frac{dJ_i}{dt}&=&\epsilon_{iab}\Big[\frac{2}{5}U_{aj}U^{(1)}_{bj}+\frac{\eta^2}{63}U_{ajk}U^{(1)}_{bjk}\nonumber\\
&+&\frac{\eta^4}{2268}U_{ajkl}U^{(1)}_{bjkl}+O\left(\eta^6\right)\Big]\,,
\eea
is identically zero in this case (because of the radial character of the motion). 

Finally, the losses of linear momentum and the position of the CM are defined as
\bea
\frac{dP_i}{dt}&=&\eta^2\Big[\frac{2}{63}U^{(1)}_{ijk}U^{(1)}_{jk}{+}\frac{\eta^2}{1134}U^{(1)}_{ijkl}U^{(1)}_{jkl}{+}O\left(\eta^6\right)\Big]\,,\nonumber\\
\frac{dG_i}{dt}&=&\mathcal{F}^i_{\rm G}=\eta^2\Big[\frac{2}{21}U_{ijk}U^{(1)}_{jk}+\eta^2\frac{2}{567}U_{ijkl}U^{(1)}_{jkl}\Big]\nonumber\\
&+&+O\left(\eta^6\right)\Big]\,,
\eea
having already considered the present simplification $V_{ij\dots}=0$. Plugging the relative orbit in these equations, we obtain
\bea
\frac{dP_1}{dt}&=&\frac{2^{2/3} \eta ^4 \nu ^2\Delta}{3^{1/3}
   T^{13/3}}\left(\frac{640 \nu
   }{15309}-\frac{2720}{15309}\right)+O\left(\eta^6\right)\,,\nonumber\\
 \frac{dP_2}{dt}&=&\frac{dP_3}{dt}=0\,,\nonumber\\
 \mathcal{F}_{\rm G}^1&=&\frac{2^{2/3} \eta ^4 \nu ^2 M\Delta  }{3^{1/3}
   T^{10/3}}\left(\frac{544}{1701}-\frac{128 \nu
   }{1701}\right)+O\left(\eta^6\right)\,,\nonumber\\
  \mathcal{F}_{\rm G}^2&=&\mathcal{F}_{\rm G}^3=0\,.\nonumber\\
\eea

In Fig. \ref{fig:2} we have shown the behavior of $\frac{dE}{d\omega}$ as a function of $\omega$ for the following choice of parameter:  $M=1$,  $\nu=[0\, \hbox{(blu online)},\frac18\, \hbox{(yellow online)},\frac14\, \hbox{(green online)}]$.
The curves show a local maximum at $[0.295142, 0.278813, 0.263673]$, respectively. 
Notice that in Refs.\cite{Davis:1971gg,Mitsou:2010jv} the radial infall of a particle is studied using the fact that only the so-called polar modes are excited, leaving a single master equation, namely the Zerilli equation. The energy spectrum in the frequency domain, obtained numerically, exhibits a maximum at $\omega M \sim 0.32$. In the Teukolsky formalism, on the other hand, the maximum is reached at $M\omega \sim 0.36$ \cite{Detweiler:1979xr}.

We note in passing that the explicit expression of the memory term is
\beq
 U_{ij}^{{\rm 2.5PN} \rm  (mem)}(T) =M\nu^2\frac{128\cdot 2^{2/3}}{3969\cdot 3^{1/3}T^{7/3}}+O(\eta^3)\,.
\eeq

\section{Inertial forces in the CM frame: effects at 4.5PN} 

The fact that the system loses linear momentum starting from the 3.5PN level of accuracy implies that the CM frame is no more an inertial frame and one has to face with the presence of inertial forces, i.e., with dragging effects. For example, if one uses a Hamiltonian description for the conservative part of the problem and adds external forces to represent both radiation-reaction and dragging effects due to inertial forces, ${\mathcal F}_i={\mathcal F}_i^{\rm rr}+{\mathcal F}_i^{\rm drag}$, the dynamics of the system is described as
\beq
\label{eq_of_moto}
\dot x^i =\frac{\partial H(x,p)}{\partial p_i}\,,\qquad \dot p_i=-\frac{\partial H(x,p)}{\partial x^i}+{\mathcal F}_i\,.
\eeq
Ref. \cite{Blanchet:2026suq} has provided us with the harmonic coordinate expression of the inertial forces which accelerate the CM (see Eq. (6.18) and (6.19) there).
Even if the latter forces start affecting the two-bodies dynamics only at 4.5PN, it is however possible to compute them since now, in order to prepare \lq\lq computational blocks" for future, highly PN-accurate, computations. Specifically, we will evaluate below the \lq\lq rad-term" of Eq. (6.10) in Ref \cite{Blanchet:2026suq} as an example. To this aim, let us define the integrated flux of the linear momentum  as
\bea
\Pi^i(t)&=&\int_{-\infty}^t dt' \frac{dP_i(t')}{dt'}\nonumber\\
&{=}&\frac{2^{2/3}\eta^4\nu^2M\Delta}{3^{1/3}T^{10/3}}\left(\frac{272}{5103}{-}\frac{64 \nu
   }{5103}\right)+O\left(\eta^6\right)\,.\nonumber\\
\eea
The radiation reaction contribution to the relative acceleration in the CM frame at 4.5PN \cite{Blanchet:2026suq} is given by
\beq\label{arr}
a_{\rm RR\, 4.5PN\, rad}^i=\frac{\eta^2\Delta}{r^3}\left(2x^iv^j+x^j  v^i\right)\Big[\Pi^j+\mathcal{F}^j_{\rm G}\Big]\,.
\eeq
Since in the present case
\beq
x^i=r (1,0,0),\quad v^i=(\dot r,0,0)=\frac{\dot r}{r} x^i\,,
\eeq
Eq. \eqref{arr} simplifies as
\bea
a_{\rm RR\, 4.5PN\, rad}^i&=&\frac{\eta^2\Delta}{r^2}\left(\mathfrak{a}^i_\Pi+\mathfrak{a}^i_{\rm G}\right)\,,\nonumber\\
\eea
with
\bea
\mathfrak{a}^i_\Pi&=& 
=3\frac{\dot r}{r^2} (x\cdot\Pi) x^i\,, \nonumber\\
\mathfrak{a}^i_{\rm G}&{=}&
3\frac{\dot r}{r^2} (x\cdot\mathcal{F}_{\rm G}) x^i\,.
\eea
A direct evaluation gives
\bea
a_{\rm RR\, 4.5PN\, rad}^1&{=}&\frac{\nu^2\eta^6}{M T^5}\left(\frac{1024 \nu
   ^2}{2187}{-}\frac{512 \nu
   }{243}{+}\frac{1088}{2187}\right)\nonumber\\
   &{+}&O\left(\eta^8\right)\,.
\eea
For completeness, we show the waveform as function or the radial coordinate. The equation $r=r_p(t)$, Eq. \eqref{orbit}, can be inverted in PN sense as follows
\bea
T_{\rm orb}(r)&=&\frac{t_{\rm orb}(r)}{M}=\frac{\sqrt{2}}{3}\left(\frac{r}{M}\right)^{3/2}\nonumber\\
&-&\sqrt{2} \eta ^2 \left(\frac{5 \nu
   }{4}-\frac{5}{2}\right)
   \sqrt{\frac{r}{M}}\nonumber\\
   &+&\sqrt{2} \eta ^4
   \sqrt{\frac{M}{r}}\left(\frac{5 \nu ^2}{32}-\frac{3 \nu
   }{4}-\frac{23}{8}\right)\nonumber\\
   &-&\frac{32 \eta ^5 \nu  M}{21 r}+O\left(\eta^6\right)\,.
\eea
Indeed Eq. \eqref{W_deco} becomes
\begin{widetext}
\bea
W(r,\theta,\phi)&=&\nu M(c_\theta c_\phi+is_\phi)^2\Bigg\{\frac{M}{2 r}+\frac{\eta^2M^2}{r^2}\cdot\frac{  (5-15 \nu )
   c_{2 \theta }+10 (3 \nu -1)
   c_{2 \phi } s_{\theta }^2+39
   \nu -9}{48 }\nonumber\\
   &+&\frac{\eta^3M^{5/2}}{r^{5/2}}\Big[\frac{ \left(\Delta 
   c_{\phi } s_{\theta } \left(7
   (2 \nu -1) c_{2 \theta }+14
   (1-2 \nu ) c_{2 \phi }
   s_{\theta }^2+58 \nu
   -229\right)+16 \sqrt{3} \pi
   +8\right)}{48 \sqrt{2} }+\sqrt{2} \mathcal{L} \Big]\nonumber\\
   &+&\frac{\eta^4M^3}{r^3}\cdot\frac{75 \nu ^2+51 \nu
   -222-25 (\nu -2)
   (3 \nu -1) c_{2 \theta }+50
   (\nu -2) (3 \nu -1) c_{2 \phi }
   s_{\theta }^2}{48 }\nonumber\\
   &+&\frac{\eta^5M^{7/2}}{r^{7/2}}\Big[\frac{ \mathcal{L}
   \left((5-15 \nu ) c_{2 \theta
   }+10 (3 \nu -1) c_{2 \phi }
   s_{\theta }^2+54 \nu
   -39\right)}{6 \sqrt{2} }\nonumber\\
   &+&\frac{1}{733824\sqrt{2}}\Big(  \left(10192 \Delta 
   (\nu  (420 \nu -2269)+1718)
   c_{\phi } s_{\theta }+31850
   \Delta  (\nu -2) (2 \nu -1)
   c_{\phi } s_{\theta } \left(-3
   c_{2 \theta }\right.\right.\nonumber\\
   &+&\left.\left.6 c_{2 \phi }
   s_{\theta }^2-1\right)+49
   \Delta  (\nu -1) (3 \nu -1)
   c_{\phi } s_{\theta } \left(44
   c_{2 \theta }+429 c_{4 \theta
   }+1144 c_{4 \phi } s_{\theta
   }^4-176 \left(13 c_{2 \theta
   }+7\right) c_{2 \phi }
   s_{\theta }^2+487\right)\right.\nonumber\\
   &-&\left.333216
   \nu  \left(7 c_{2 \theta }-14
   c_{2 \phi } s_{\theta
   }^2-3\right)+14560 \sqrt{3} \pi
    (3 \nu -1) \left(7 c_{2 \theta
   }-14 c_{2 \phi } s_{\theta
   }^2-3\right)\right.\nonumber\\
   &+&\left.112112 \left(7
   c_{2 \theta }-14 c_{2 \phi }
   s_{\theta }^2-3\right)-6394128
   \nu +8736 \sqrt{3} \pi  (239
   \nu -614)+30759456\right)\Big)\Big]\Bigg\}+O\left(\eta^6\right)
\eea
\end{widetext}
\section{Concluding remarks}

We have applied the Multipolar Post Minkowskian formalism to the case of a two-body system in radial fall (or a head-on collision situation). We have computed, within the post-Newtonian (PN) approximation, the corresponding waveform, up to the 2.5 PN accuracy level. At this level the presence of a radiation-reaction force manifests,  modifying the fall with a corresponding bremsstrahlung radiation. We have evaluated  all the emissions: energy, angular momentum (vanishing identically) and linear momentum. Finally, we have also evaluated the (nonlocal)   inertial forces contributions (appearing at 4.5PN) in the CM due to all these losses, paving the way for future more accurate computations (we point out that in principle, for the waveform, one can reach only the 4PN accuracy by using existing literature, see e.g. Ref. \cite{Blanchet:2013haa}). An associated ancillary file contains the explicit expressions for all the main accomplishments of the present work \cite{anc}.

As already stated in the Introduction, the PN expansion for a radially infalling particle does not provide access to the full solution of the problem. Indeed, by definition, the PN expansion can reconstruct the relevant observables only in the weak field regime, i.e., far from the BH horizon where the radial fall terminates and where the particle inevitably probes the strong-field region. For this reason, the present work is also preliminary to future, forthcoming analytical studies of the same problem directly in the strong-field regime.

The leading contribution to the power emitted at infinity in the limit $GM\omega \gg 1$ (not accessible with the  PN treatment considered here, as already stated) was computed in Ref.\cite{Zerilli:1970wzz} and exhibits an exponentially decreasing behavior, $e^{-4\pi M\omega}$. It would be interesting to develop a systematic procedure to compute sub-leading corrections in the large-$GM\omega$ expansion. The ultimate goal would be to match the weak and strong field results in their overlapping region.

Besides the PN approach developed here, another context in which this type of physics, related to the GWs emission, naturally arises is the scattering amplitudes one (see e.g. the last accomplishment at the 1 loop level \cite{Brandhuber:2023hhy,Herderschee:2023fxh,Georgoudis:2023lgf}. However, the specific problem of radial infall is not easily treated neither at the tree level from the amplitudes, since it is not fully clear yet how to identify a suitable expansion parameter along the falling process. 
Several attempts are currently under active consideration. 
For example, one can study a non radial falling process, in which a particle is released from infinity with a small but non vanishing angular momentum (or equivalently, with a small impact parameter) with respect to the critical value associated with  capture by the BH:  $b/b_c \ll 1$, where $b_c = 3\sqrt{3}M$ corresponds to the capture \cite{Bianchi:2021yqs,Bianchi:2022wku,Cipriani:2025ini}.

Finally, this problem can be framed into  a self-force perspective too. As usual, a synergic effort by all these different approach can be a keystone to reach a satisfying solution of this problem.

\section*{Acknowledgments}

D.B. and G.D.R. thank S. Albanesi, A. Cipriani, S. de Angelis, D. Fioravanti, A. Geralico, A. Nagar, R. Roiban and  M. Rossi for fruitful discussions.
D.B. and G.D.R. acknowledge  membership to the Italian Gruppo Nazionale per
la Fisica Matematica (GNFM) of the Istituto Nazionale
di Alta Matematica (INDAM). D.B. thanks the hospitality
and the highly stimulating environment of the Institut
des Hautes Etudes Scient\'if\'iques (FR) since this work was completed while visiting it.

\begin{widetext}
\appendix

\section{Decomposing the waveform in spin-weighted spherical harmonics}

Here we list the mode-decomposition on the spin-weighted spherical harmonics of the various radiative multipoles.
The corresponding expression for the asymptotic metric mode decomposition is given in Eq. (187) of Ref.\cite{Blanchet:2013haa} and reads
\beq
h_{\ell m}=-\frac{G}{\sqrt{2}r c^{\ell+2}}U_{\ell m}
\eeq
also largely used in literature (see e.g. \cite{Lousto:1996sx,Lousto:1997wf,Martel:2001yf,Nagar:2006xv}).

\begin{table*}[h!]
\caption{\label{decY}Decomposition of the waveform in the basis of the spin-weighted spherical harmonics.}
\begin{ruledtabular}
\begin{tabular}{lll}
& $U_{2,\pm2}$ & ${-}\frac{ ({-}2)^{1/3} \sqrt{\pi }
   \nu  M^2}{3^{1/6}
   \sqrt{5}}\Bigg[\frac{\Gamma
   \left(\frac{1}{3}\right)}{ (M\omega)^{1/3}}{-}\frac{1}{7} ({-}6)^{1/3} \eta ^2
   (8 \nu {-}33) 
   (M\omega)^{1/3} \Gamma
   \left(\frac{2}{3}\right){+}\frac{i}{6}\eta^3\Gamma\left(\frac{1}{3}\right)(M\omega)^{2/3}(12(\gamma{+}\mathcal{L}{+}\frac{i\pi}{2}){-}17{+}18\log(3))$\\
  &  &${+}\frac{\eta^5({-}1)^{5/6} (M \omega)^{4/3}}{49\cdot 6^{2/3}}
   \Gamma
   \left(\frac{2}{3}\right)
   (84 \mathcal{L} (33{-}8 \nu ){+}84 \gamma 
   (33{-}8 \nu ){+}862 \nu {-}42 i \pi
    (8 \nu {-}33){+}126 (33{-}8 \nu )
   \log (3){-}3927)\Big]$\\ 
$U_2$& $U_{2,\pm1}$ & $0$\\
 & $U_{2,0}$ & $-\sqrt{\frac{2}{3}}U_{2,2}$\\
\hline
& $U_{3,\pm3}$ & $\pm\frac{(-1)^{5/6} \nu \sqrt{\pi} \sqrt{1
   -4  \nu } M^2}{9
   \cdot3^{2/3}\cdot2^{1/6}  \sqrt{7}}\Big[\frac{5}{2} \eta ^2 (4 \nu -17) 
   (M\omega)^{2/3} \Gamma
   \left(\frac{1}{3}\right)-\frac{(-3)^{1/3} \eta ^4 (20\nu^2-214\nu+633) 
   (M\omega)^{4/3} \Gamma
   \left(\frac{2}{3}\right)}{2\cdot
   2^{2/3}}\Big]$\\
 & $U_{3,\pm 2}$ & $0$\\
$U_3$ & $U_{3,\pm1}$ & $\mp\sqrt{\frac{3}{5}}U_{3,3}$\\
  & $U_{3,0}$ & $0$\\
\hline
 & $U_{4,\pm4}$ & ${-}\frac{5(-1)^{2/3} \sqrt{\frac{\pi }{7}}
   \nu  (3 \nu {-}1) M^2}{6\cdot
   3^{5/6}2^{1/3} } (M\omega)^{1/3}
   \Gamma
   \left(\frac{2}{3}\right)$\\
   & $U_{4,\pm3}$ & $0$\\
   & $U_{4,\pm2}$ & $-\frac{2}{\sqrt{7}}U_{4,4}$\\
  $U_4$   & $U_{4,\pm1}$ & $0$\\
 & $U_{4,0}$ & $3\sqrt{\frac{2}{35}}U_{4,4}$\\
   \hline
   & $U_{5,\pm5}$ & $\mp\frac{7 (-1)^{5/6} \nu  (2 \nu
   -1) \sqrt{\pi}\sqrt{1 -4   \nu }
   M^2}{24\cdot 3^{2/3} \cdot 2^{1/6} 
   \sqrt{55}}\Big[(M \omega)^{2/3} \Gamma
   \left(\frac{1}{3}\right)-\frac{25}{7} (-6)^{1/3} \eta ^2
   (\nu -2) (M \omega)^{4/3}
   \Gamma
   \left(\frac{2}{3}\right)\Big]$\\
   & $U_{5,\pm4}$ & $0$\\
   & $U_{5,\pm3}$ & $\mp\frac{\sqrt{5}}{3}U_{5,5}$\\
   & $U_{5,\pm2}$ & $0$\\
 $U_5$  & $U_{5,\pm1}$ & $\pm\sqrt{\frac{10}{21}}U_{5,5}$\\
   & $U_{5,0}$ & $0$\\
\hline
   & $U_{7,\pm7}$ & $\mp\frac{(-1)^{1/6}
   \sqrt{\frac{11 \pi }{455}}
   \sqrt{1-4 \nu } (\nu -1) \nu 
   (3 \nu -1) M^{10/3} \omega
   ^{4/3} \Gamma
   \left(\frac{2}{3}\right)}{18\cdot
   2^{5/6} 3^{1/3}}$\\
   & $U_{7,\pm6}$ & $0$\\
   & $U_{7,\pm5}$ & $\mp\sqrt{\frac{7}{13}}U_{7,7}$\\
   & $U_{7,\pm4}$ & $0$\\
   & $U_{7,\pm3}$ & $\pm3\sqrt{\frac{7}{143}}U_{7,7}$\\
$U_7$   & $U_{7,\pm2}$ & $0$\\
   & $U_{7,\pm1}$ & $\mp5\sqrt{\frac{7}{429}}U_{7,7}$\\
   & $U_{7,0}$ & $0$\\
\end{tabular}
\end{ruledtabular}
\end{table*}
\end{widetext}
\FloatBarrier
\bibliographystyle{apsrev}
\bibliography{references}

@article{Bini:2023fiz,
    author = "Bini, Donato and Damour, Thibault and Geralico, Andrea",
    title = "{Comparing one-loop gravitational bremsstrahlung amplitudes to the multipolar-post-Minkowskian waveform}",
    eprint = "2309.14925",
    archivePrefix = "arXiv",
    primaryClass = "gr-qc",
    doi = "10.1103/PhysRevD.108.124052",
    journal = "Phys. Rev. D",
    volume = "108",
    number = "12",
    pages = "124052",
    year = "2023"
}

@article{Martel:2001yf,
    author = "Martel, Karl and Poisson, Eric",
    title = "{A One parameter family of time symmetric initial data for the radial infall of a particle into a Schwarzschild black hole}",
    eprint = "gr-qc/0107104",
    archivePrefix = "arXiv",
    doi = "10.1103/PhysRevD.66.084001",
    journal = "Phys. Rev. D",
    volume = "66",
    pages = "084001",
    year = "2002"
}

@article{Nagar:2006xv,
    author = "Nagar, Alessandro and Damour, Thibault and Tartaglia, Angelo",
    editor = "Campanelli, Manuela and Rezzolla, Luciano",
    title = "{Binary black hole merger in the extreme mass ratio limit}",
    eprint = "gr-qc/0612096",
    archivePrefix = "arXiv",
    doi = "10.1088/0264-9381/24/12/S08",
    journal = "Class. Quant. Grav.",
    volume = "24",
    pages = "S109--S124",
    year = "2007"
}

@article{Lousto:1996sx,
    author = "Lousto, Carlos O. and Price, Richard H.",
    title = "{Headon collisions of black holes: The Particle limit}",
    eprint = "gr-qc/9609012",
    archivePrefix = "arXiv",
    doi = "10.1103/PhysRevD.55.2124",
    journal = "Phys. Rev. D",
    volume = "55",
    pages = "2124--2138",
    year = "1997"
}

@article{Lousto:1997wf,
    author = "Lousto, Carlos O. and Price, Richard H.",
    title = "{Understanding initial data for black hole collisions}",
    eprint = "gr-qc/9705071",
    archivePrefix = "arXiv",
    doi = "10.1103/PhysRevD.56.6439",
    journal = "Phys. Rev. D",
    volume = "56",
    pages = "6439--6457",
    year = "1997"
}

@article{Scheel:2025jct,
    author = "Scheel, Mark A. and others",
    title = "{The SXS collaboration{\textquoteright}s third catalog of binary black hole simulations}",
    eprint = "2505.13378",
    archivePrefix = "arXiv",
    primaryClass = "gr-qc",
    doi = "10.1088/1361-6382/adfd34",
    journal = "Class. Quant. Grav.",
    volume = "42",
    number = "19",
    pages = "195017",
    year = "2025"
}

@article{Brandhuber:2023hhy,
    author = "Brandhuber, Andreas and Brown, Graham R. and Chen, Gang and De Angelis, Stefano and Gowdy, Joshua and Travaglini, Gabriele",
    title = "{One-loop gravitational bremsstrahlung and waveforms from a heavy-mass effective field theory}",
    eprint = "2303.06111",
    archivePrefix = "arXiv",
    primaryClass = "hep-th",
    reportNumber = "QMUL-22-28,SAGEX-22-32-E",
    doi = "10.1007/JHEP06(2023)048",
    journal = "JHEP",
    volume = "06",
    pages = "048",
    year = "2023"
}

@article{DeAmicis:2024eoy,
    author = "De Amicis, Marina and others",
    title = "{Late-Time Tails in Nonlinear Evolutions of Merging Black Holes}",
    eprint = "2412.06887",
    archivePrefix = "arXiv",
    primaryClass = "gr-qc",
    doi = "10.1103/2brx-xnyr",
    journal = "Phys. Rev. Lett.",
    volume = "135",
    number = "17",
    pages = "171401",
    year = "2025"
}

@article{Georgoudis:2023lgf,
    author = "Georgoudis, Alessandro and Heissenberg, Carlo and Vazquez-Holm, Ingrid",
    title = "{Inelastic exponentiation and classical gravitational scattering at one loop}",
    eprint = "2303.07006",
    archivePrefix = "arXiv",
    primaryClass = "hep-th",
    reportNumber = "NORDITA 2023-010, UUITP-03/23",
    doi = "10.1007/JHEP06(2023)126",
    journal = "JHEP",
    volume = "2023",
    number = "06",
    pages = "126",
    year = "2023"
}

@article{Herderschee:2023fxh,
    author = "Herderschee, Aidan and Roiban, Radu and Teng, Fei",
    title = "{The sub-leading scattering waveform from amplitudes}",
    eprint = "2303.06112",
    archivePrefix = "arXiv",
    primaryClass = "hep-th",
    reportNumber = "LCTP-23-04",
    doi = "10.1007/JHEP06(2023)004",
    journal = "JHEP",
    volume = "06",
    pages = "004",
    year = "2023"
}

@article{Bianchi:2021yqs,
    author = "Bianchi, Massimo and Di Russo, Giorgio",
    title = "{Turning black holes and D-branes inside out of their photon spheres}",
    eprint = "2110.09579",
    archivePrefix = "arXiv",
    primaryClass = "hep-th",
    doi = "10.1103/PhysRevD.105.126007",
    journal = "Phys. Rev. D",
    volume = "105",
    number = "12",
    pages = "126007",
    year = "2022"
}

@article{Cipriani:2025ini,
    author = "Cipriani, Andrea and De Santis, Alessandro and Di Russo, Giorgio and Grillo, Alfredo and Tabarroni, Luca",
    title = "{Hamiltonian neural network approach to fuzzball geodesics}",
    eprint = "2502.20881",
    archivePrefix = "arXiv",
    primaryClass = "hep-th",
    doi = "10.1103/dssv-x49b",
    journal = "Phys. Rev. D",
    volume = "112",
    number = "2",
    pages = "026018",
    year = "2025"
}

@article{Bianchi:2022wku,
    author = "Bianchi, Massimo and Di Russo, Giorgio",
    title = "{Turning rotating D-branes and black holes inside out their photon-halo}",
    eprint = "2203.14900",
    archivePrefix = "arXiv",
    primaryClass = "hep-th",
    doi = "10.1103/PhysRevD.106.086009",
    journal = "Phys. Rev. D",
    volume = "106",
    number = "8",
    pages = "086009",
    year = "2022"
}

@article{Cipriani:2026xmx,
    author = "Cipriani, Andrea and Nagar, Alessandro and Fucito, Francesco and Morales, Jos{\'e} Francisco",
    title = "{From the confluent Heun equation to a new factorized and resummed gravitational waveform for circularized, nonspinning, compact binaries}",
    eprint = "2602.08833",
    archivePrefix = "arXiv",
    primaryClass = "gr-qc",
    month = "2",
    year = "2026"
}

@article{Cipriani:2026myb,
    author = "Cipriani, Andrea and Fucito, Francesco and Heissenberg, Carlo and Morales, Jose Francisco and Russo, Rodolfo",
    title = "{''Waveforms'' at the Horizon}",
    eprint = "2602.05766",
    archivePrefix = "arXiv",
    primaryClass = "gr-qc",
    month = "2",
    year = "2026"
}

@article{Damour:1981bh,
    author = "Damour, Thibault and Deruelle, Nathalie",
    title = "{Radiation Reaction and Angular Momentum Loss in Small Angle Gravitational Scattering}",
    reportNumber = "PRINT-81-0533 (MEUDON)",
    doi = "10.1016/0375-9601(81)90567-3",
    journal = "Phys. Lett. A",
    volume = "87",
    pages = "81",
    year = "1981"
}

@article{Zerilli:1970se,
    author = "Zerilli, Frank J.",
    title = "{Effective potential for even parity Regge-Wheeler gravitational perturbation equations}",
    doi = "10.1103/PhysRevLett.24.737",
    journal = "Phys. Rev. Lett.",
    volume = "24",
    pages = "737--738",
    year = "1970"
}

@article{Bini:2024rsy,
    author = "Bini, Donato and Damour, Thibault and De Angelis, Stefano and Geralico, Andrea and Herderschee, Aidan and Roiban, Radu and Teng, Fei",
    title = "{Gravitational waveforms: A tale of two formalisms}",
    eprint = "2402.06604",
    archivePrefix = "arXiv",
    primaryClass = "hep-th",
    doi = "10.1103/PhysRevD.109.125008",
    journal = "Phys. Rev. D",
    volume = "109",
    number = "12",
    pages = "125008",
    year = "2024"
}

@article{Bini:2024ijq,
    author = "Bini, Donato and Damour, Thibault and Geralico, Andrea",
    title = "{Gravitational bremsstrahlung waveform at the fourth post-Minkowskian order and the second post-Newtonian level}",
    eprint = "2407.02076",
    archivePrefix = "arXiv",
    primaryClass = "gr-qc",
    doi = "10.1103/PhysRevD.110.064035",
    journal = "Phys. Rev. D",
    volume = "110",
    number = "6",
    pages = "064035",
    year = "2024"
}

@article{Blanchet:2013haa,
    author = "Blanchet, Luc",
    title = "{Post-Newtonian Theory for Gravitational Waves}",
    eprint = "1310.1528",
    archivePrefix = "arXiv",
    primaryClass = "gr-qc",
    doi = "10.12942/lrr-2014-2",
    journal = "Living Rev. Rel.",
    volume = "17",
    pages = "2",
    year = "2014"
}

@article{Bini:2026cpw,
    author = "Bini, Donato and Di Russo, Giorgio",
    title = "{Analytic self-force effects on radial infalling particles in the Schwarzschild spacetime: the radiated energy}",
    eprint = "2601.11186",
    archivePrefix = "arXiv",
    primaryClass = "gr-qc",
    month = "1",
    year = "2026"
}

@book{Maggiore:2007ulw,
    author = "Maggiore, Michele",
    title = "{Gravitational Waves. Vol. 1: Theory and Experiments}",
    doi = "10.1093/acprof:oso/9780198570745.001.0001",
    isbn = "978-0-19-171766-6, 978-0-19-852074-0",
    publisher = "Oxford University Press",
    year = "2007"
}

@article{Arun:2007sg,
    author = "Arun, K. G. and Blanchet, Luc and Iyer, Bala R. and Qusailah, Moh'd S. S.",
    title = "{Inspiralling compact binaries in quasi-elliptical orbits: The Complete 3PN energy flux}",
    eprint = "0711.0302",
    archivePrefix = "arXiv",
    primaryClass = "gr-qc",
    doi = "10.1103/PhysRevD.77.064035",
    journal = "Phys. Rev. D",
    volume = "77",
    pages = "064035",
    year = "2008"
}

@article{Blanchet:2026suq,
    author = "Blanchet, Luc and Faye, Guillaume and Seraille, Emeric and Trestini, David",
    title = "{Gravitational radiation reaction for compact binary systems at the fourth-and-a-half post-Newtonian order in harmonic coordinates}",
    eprint = "2601.06743",
    archivePrefix = "arXiv",
    primaryClass = "gr-qc",
    month = "1",
    year = "2026"
}

@article{Poisson:2003nc,
    author = "Poisson, Eric",
    title = "{The Motion of point particles in curved space-time}",
    eprint = "gr-qc/0306052",
    archivePrefix = "arXiv",
    doi = "10.12942/lrr-2004-6",
    journal = "Living Rev. Rel.",
    volume = "7",
    pages = "6",
    year = "2004"
}

@article{Barack:2018yvs,
    author = "Barack, Leor and Pound, Adam",
    title = "{Self-force and radiation reaction in general relativity}",
    eprint = "1805.10385",
    archivePrefix = "arXiv",
    primaryClass = "gr-qc",
    doi = "10.1088/1361-6633/aae552",
    journal = "Rept. Prog. Phys.",
    volume = "82",
    number = "1",
    pages = "016904",
    year = "2019"
}

@article{Zerilli:1970wzz,
    author = "Zerilli, F. J.",
    title = "{Gravitational field of a particle falling in a schwarzschild geometry analyzed in tensor harmonics}",
    doi = "10.1103/PhysRevD.2.2141",
    journal = "Phys. Rev. D",
    volume = "2",
    pages = "2141--2160",
    year = "1970"
}

@article{Davis:1971gg,
    author = "Davis, M. and Ruffini, R. and Press, W. H. and Price, R. H.",
    title = "{Gravitational radiation from a particle falling radially into a schwarzschild black hole}",
    doi = "10.1103/PhysRevLett.27.1466",
    journal = "Phys. Rev. Lett.",
    volume = "27",
    pages = "1466--1469",
    year = "1971"
}

@article{Davis:1972ud,
    author = "Davis, M. and Ruffini, R. and Tiomno, J.",
    title = "{Pulses of gravitational radiation of a particle falling radially into a schwarzschild black hole}",
    doi = "10.1103/PhysRevD.5.2932",
    journal = "Phys. Rev. D",
    volume = "5",
    pages = "2932--2935",
    year = "1972"
}

@article{Ruffini:1973ky,
    author = "Ruffini, R.",
    title = "{Gravitational radiation from a mass projected into a schwarzschild black hole}",
    doi = "10.1103/PhysRevD.7.972",
    journal = "Phys. Rev. D",
    volume = "7",
    pages = "972--976",
    year = "1973"
}

@article{Mitsou:2010jv,
    author = "Mitsou, Ermis",
    title = "{Gravitational radiation from radial infall of a particle into a Schwarzschild black hole. A numerical study of the spectra, quasi-normal modes and power-law tails}",
    eprint = "1012.2028",
    archivePrefix = "arXiv",
    primaryClass = "gr-qc",
    doi = "10.1103/PhysRevD.83.044039",
    journal = "Phys. Rev. D",
    volume = "83",
    pages = "044039",
    year = "2011"
}

@article{Oliveira:2018pnx,
    author = "Oliveira, Leandro A. and Crispino, Lu{\'\i}s C. B. and Higuchi, Atsushi",
    title = "{Scalar radiation from a radially infalling source into a Schwarzschild black hole in the framework of quantum field theory}",
    doi = "10.1140/epjc/s10052-018-5604-8",
    journal = "Eur. Phys. J. C",
    volume = "78",
    number = "2",
    pages = "133",
    year = "2018"
}

@article{Barausse:2021xes,
    author = "Barausse, Enrico and Berti, Emanuele and Cardoso, Vitor and Hughes, Scott A. and Khanna, Gaurav",
    title = "{Divergences in gravitational-wave emission and absorption from extreme mass ratio binaries}",
    eprint = "2106.09721",
    archivePrefix = "arXiv",
    primaryClass = "gr-qc",
    doi = "10.1103/PhysRevD.104.064031",
    journal = "Phys. Rev. D",
    volume = "104",
    number = "6",
    pages = "064031",
    year = "2021"
}

@article{LIGOScientific:2025rid,
    author = "Abac, A. G. and others",
    collaboration = "LIGO Scientific, Virgo, KAGRA",
    title = "{GW250114: Testing Hawking{\textquoteright}s Area Law and the Kerr Nature of Black Holes}",
    eprint = "2509.08054",
    archivePrefix = "arXiv",
    primaryClass = "gr-qc",
    reportNumber = "LIGO-P2500421",
    doi = "10.1103/kw5g-d732",
    journal = "Phys. Rev. Lett.",
    volume = "135",
    number = "11",
    pages = "111403",
    year = "2025"
}

@article{Chandrasekhar:1975zza,
    author = "Chandrasekhar, S. and Detweiler, Steven L.",
    title = "{The quasi-normal modes of the Schwarzschild black hole}",
    doi = "10.1098/rspa.1975.0112",
    journal = "Proc. Roy. Soc. Lond. A",
    volume = "344",
    pages = "441--452",
    year = "1975"
}

@article{Berti:2009kk,
    author = "Berti, Emanuele and Cardoso, Vitor and Starinets, Andrei O.",
    title = "{Quasinormal modes of black holes and black branes}",
    eprint = "0905.2975",
    archivePrefix = "arXiv",
    primaryClass = "gr-qc",
    doi = "10.1088/0264-9381/26/16/163001",
    journal = "Class. Quant. Grav.",
    volume = "26",
    pages = "163001",
    year = "2009"
}

@article{Aminov:2020yma,
    author = "Aminov, Gleb and Grassi, Alba and Hatsuda, Yasuyuki",
    title = "{Black Hole Quasinormal Modes and Seiberg{\textendash}Witten Theory}",
    eprint = "2006.06111",
    archivePrefix = "arXiv",
    primaryClass = "hep-th",
    reportNumber = "RUP-20-18",
    doi = "10.1007/s00023-021-01137-x",
    journal = "Annales Henri Poincare",
    volume = "23",
    number = "6",
    pages = "1951--1977",
    year = "2022"
}

@article{Bianchi:2021xpr,
    author = "Bianchi, Massimo and Consoli, Dario and Grillo, Alfredo and Morales, Jos{\`e} Francisco",
    title = "{QNMs of branes, BHs and fuzzballs from quantum SW geometries}",
    eprint = "2105.04245",
    archivePrefix = "arXiv",
    primaryClass = "hep-th",
    doi = "10.1016/j.physletb.2021.136837",
    journal = "Phys. Lett. B",
    volume = "824",
    pages = "136837",
    year = "2022"
}

@article{Bianchi:2021mft,
    author = "Bianchi, Massimo and Consoli, Dario and Grillo, Alfredo and Morales, Jose Francisco",
    title = "{More on the SW-QNM correspondence}",
    eprint = "2109.09804",
    archivePrefix = "arXiv",
    primaryClass = "hep-th",
    doi = "10.1007/JHEP01(2022)024",
    journal = "JHEP",
    volume = "01",
    pages = "024",
    year = "2022"
}

@article{Bianchi:2022qph,
    author = "Bianchi, Massimo and Di Russo, Giorgio",
    title = "{2-charge circular fuzz-balls and their perturbations}",
    eprint = "2212.07504",
    archivePrefix = "arXiv",
    primaryClass = "hep-th",
    doi = "10.1007/JHEP08(2023)217",
    journal = "JHEP",
    volume = "08",
    pages = "217",
    year = "2023"
}

@article{Bianchi:2023sfs,
    author = "Bianchi, Massimo and Di Russo, Giorgio and Grillo, Alfredo and Morales, Jose Francisco and Sudano, Giuseppe",
    title = "{On the stability and deformability of top stars}",
    eprint = "2305.15105",
    archivePrefix = "arXiv",
    primaryClass = "gr-qc",
    doi = "10.1007/JHEP12(2023)121",
    journal = "JHEP",
    volume = "12",
    pages = "121",
    year = "2023"
}

@article{DiRusso:2024hmd,
    author = "Di Russo, Giorgio and Fucito, Francesco and Morales, Jose Francisco",
    title = "{Tidal resonances for fuzzballs}",
    eprint = "2402.06621",
    archivePrefix = "arXiv",
    primaryClass = "hep-th",
    doi = "10.1007/JHEP04(2024)149",
    journal = "JHEP",
    volume = "04",
    pages = "149",
    year = "2024"
}

@article{Bena:2024hoh,
    author = "Bena, Iosif and Di Russo, Giorgio and Morales, Jose Francisco and Ruip{\'e}rez, Alejandro",
    title = "{Non-spinning tops are stable}",
    eprint = "2406.19330",
    archivePrefix = "arXiv",
    primaryClass = "hep-th",
    doi = "10.1007/JHEP10(2024)071",
    journal = "JHEP",
    volume = "10",
    pages = "071",
    year = "2024"
}

@article{Cipriani:2024ygw,
    author = "Cipriani, Andrea and Di Benedetto, Carlo and Di Russo, Giorgio and Grillo, Alfredo and Sudano, Giuseppe",
    title = "{Charge (in)stability and superradiance of Topological Stars}",
    eprint = "2405.06566",
    archivePrefix = "arXiv",
    primaryClass = "hep-th",
    doi = "10.1007/JHEP07(2024)143",
    journal = "JHEP",
    volume = "07",
    pages = "143",
    year = "2024"
}

@article{Fioravanti:2021dce,
    author = "Fioravanti, Davide and Gregori, Daniele",
    title = "{New method for exact results on quasinormal modes of black holes}",
    eprint = "2112.11434",
    archivePrefix = "arXiv",
    primaryClass = "hep-th",
    doi = "10.1103/b8pl-vdwy",
    journal = "Phys. Rev. D",
    volume = "112",
    number = "12",
    pages = "125020",
    year = "2025"
}

@article{DiRusso:2025qpf,
    author = "Di Russo, Giorgio and Tokareva, Anna",
    title = {{Scalar Quasinormal modes in Reissner--Nordstr{\"o}m black holes: implications for Weak Gravity Conjecture}},
    eprint = "2510.06813",
    archivePrefix = "arXiv",
    primaryClass = "hep-th",
    month = "10",
    year = "2025"
}

@book{Maggiore:2018sht,
    author = "Maggiore, Michele",
    title = "{Gravitational Waves. Vol. 2: Astrophysics and Cosmology}",
    isbn = "978-0-19-857089-9",
    publisher = "Oxford University Press",
    month = "3",
    year = "2018"
}

@article{Teukolsky:1973ha,
    author = "Teukolsky, Saul A.",
    title = "{Perturbations of a rotating black hole. 1. Fundamental equations for gravitational electromagnetic and neutrino field perturbations}",
    doi = "10.1086/152444",
    journal = "Astrophys. J.",
    volume = "185",
    pages = "635--647",
    year = "1973"
}

@article{Press:1973zz,
    author = "Press, William H. and Teukolsky, Saul A.",
    title = "{Perturbations of a Rotating Black Hole. II. Dynamical Stability of the Kerr Metric}",
    doi = "10.1086/152445",
    journal = "Astrophys. J.",
    volume = "185",
    pages = "649--674",
    year = "1973"
}

@article{Mano:1996vt,
    author = "Mano, Shuhei and Suzuki, Hisao and Takasugi, Eiichi",
    title = "{Analytic solutions of the Teukolsky equation and their low frequency expansions}",
    eprint = "gr-qc/9603020",
    archivePrefix = "arXiv",
    reportNumber = "OU-HET-238",
    doi = "10.1143/PTP.95.1079",
    journal = "Prog. Theor. Phys.",
    volume = "95",
    pages = "1079--1096",
    year = "1996"
}

@article{Teukolsky:1974yv,
    author = "Teukolsky, S. A. and Press, W. H.",
    title = "{Perturbations of a rotating black hole. III - Interaction of the hole with gravitational and electromagnetic radiation}",
    doi = "10.1086/153180",
    journal = "Astrophys. J.",
    volume = "193",
    pages = "443--461",
    year = "1974"
}

@article{Bah:2020pdz,
    author = "Bah, Ibrahima and Heidmann, Pierre",
    title = "{Topological stars, black holes and generalized charged Weyl solutions}",
    eprint = "2012.13407",
    archivePrefix = "arXiv",
    primaryClass = "hep-th",
    doi = "10.1007/JHEP09(2021)147",
    journal = "JHEP",
    volume = "09",
    pages = "147",
    year = "2021"
}

@article{Heidmann:2025pbb,
    author = "Heidmann, Pierre and Pani, Paolo and Santos, Jorge E.",
    title = "{Asymptotically flat rotating topological stars}",
    eprint = "2510.05200",
    archivePrefix = "arXiv",
    primaryClass = "hep-th",
    doi = "10.1007/JHEP03(2026)108",
    journal = "JHEP",
    volume = "03",
    pages = "108",
    year = "2026"
}

@article{Bianchi:2025uis,
    author = "Bianchi, Massimo and Dibitetto, Giuseppe and Morales, Jose F. and Ruip{\'e}rez, Alejandro",
    title = "{Rotating Topological Stars}",
    eprint = "2504.12235",
    archivePrefix = "arXiv",
    primaryClass = "hep-th",
    doi = "10.1007/JHEP01(2026)046",
    journal = "JHEP",
    volume = "01",
    pages = "046",
    year = "2026"
}

@article{Dima:2025tjz,
    author = "Dima, Alexandru and Heidmann, Pierre and Melis, Marco and Pani, Paolo and Patashuri, Gela",
    title = "{W-solitons as prototypical black hole microstates}",
    eprint = "2509.18245",
    archivePrefix = "arXiv",
    primaryClass = "gr-qc",
    doi = "10.1103/2wcq-4xny",
    journal = "Phys. Rev. D",
    volume = "112",
    number = "12",
    pages = "124056",
    year = "2025"
}

@article{Bianchi:2024vmi,
    author = "Bianchi, Massimo and Bini, Donato and Di Russo, Giorgio",
    title = "{Scalar perturbations of topological-star spacetimes}",
    eprint = "2407.10868",
    archivePrefix = "arXiv",
    primaryClass = "gr-qc",
    doi = "10.1103/PhysRevD.110.084077",
    journal = "Phys. Rev. D",
    volume = "110",
    number = "8",
    pages = "084077",
    year = "2024"
}

@article{Ivanov:2025ozg,
    author = "Ivanov, Mikhail M. and Li, Yue-Zhou and Parra-Martinez, Julio and Zhou, Zihan",
    title = "{Resummation of Universal Tails in Gravitational Waveforms}",
    eprint = "2504.07862",
    archivePrefix = "arXiv",
    primaryClass = "hep-th",
    reportNumber = "MIT-CTP/5863",
    doi = "10.1103/jzd1-qzkt",
    journal = "Phys. Rev. Lett.",
    volume = "135",
    number = "14",
    pages = "141401",
    year = "2025"
}

@article{Bianchi:2024rod,
    author = "Bianchi, Massimo and Bini, Donato and Di Russo, Giorgio",
    title = "{Scalar waves in a topological star spacetime: Self-force and radiative losses}",
    eprint = "2411.19612",
    archivePrefix = "arXiv",
    primaryClass = "gr-qc",
    doi = "10.1103/PhysRevD.111.044017",
    journal = "Phys. Rev. D",
    volume = "111",
    number = "4",
    pages = "044017",
    year = "2025"
}

@article{Fucito:2024wlg,
    author = "Fucito, Francesco and Morales, Jose Francisco and Russo, Rodolfo",
    title = "{Gravitational wave forms for extreme mass ratio collisions from supersymmetric gauge theories}",
    eprint = "2408.07329",
    archivePrefix = "arXiv",
    primaryClass = "hep-th",
    doi = "10.1103/PhysRevD.111.044054",
    journal = "Phys. Rev. D",
    volume = "111",
    number = "4",
    pages = "044054",
    year = "2025"
}

@article{DiRusso:2025lip,
    author = "Di Russo, Giorgio and Bianchi, Massimo and Bini, Donato",
    title = "{Scalar waves from unbound orbits in a topological star spacetime: PN reconstruction of the field and radiation losses in a self-force approach}",
    eprint = "2502.21040",
    archivePrefix = "arXiv",
    primaryClass = "gr-qc",
    doi = "10.1103/sycf-brn1",
    journal = "Phys. Rev. D",
    volume = "112",
    number = "2",
    pages = "024002",
    year = "2025"
}

@article{Bianchi:2025aei,
    author = "Bianchi, Massimo and Bini, Donato and Di Russo, Giorgio",
    title = "{Scattering angle in a topological star spacetime: A self-force approach}",
    eprint = "2506.04876",
    archivePrefix = "arXiv",
    primaryClass = "gr-qc",
    doi = "10.1103/cz6k-wqpn",
    journal = "Phys. Rev. D",
    volume = "112",
    number = "4",
    pages = "044008",
    year = "2025"
}

@article{Bini:2025ltr,
    author = "Bini, Donato and Di Russo, Giorgio",
    title = "{Topological stars and scalar wave equation: Exact resummation of the renormalized angular momentum in the eikonal limit}",
    eprint = "2506.14442",
    archivePrefix = "arXiv",
    primaryClass = "gr-qc",
    doi = "10.1103/dw5y-4pv8",
    journal = "Phys. Rev. D",
    volume = "112",
    number = "6",
    pages = "064008",
    year = "2025"
}

@article{Bianchi:2025ydq,
    author = "Bianchi, Massimo and Bini, Donato and Di Russo, Giorgio",
    title = "{Scalar self-force effects in neutral W-soliton backgrounds}",
    eprint = "2511.01402",
    archivePrefix = "arXiv",
    primaryClass = "gr-qc",
    doi = "10.1103/rf8g-yl2d",
    journal = "Phys. Rev. D",
    volume = "113",
    number = "4",
    pages = "044028",
    year = "2026"
}

@article{Bini:2025bll,
    author = "Bini, Donato and Di Russo, Giorgio and Geralico, Andrea",
    title = "{Kerr spacetime and scalar wave equation: Exact resummation of the renormalized angular momentum in the eikonal limit}",
    eprint = "2508.12046",
    archivePrefix = "arXiv",
    primaryClass = "gr-qc",
    doi = "10.1103/mzqw-wbvf",
    journal = "Phys. Rev. D",
    volume = "112",
    number = "6",
    pages = "064077",
    year = "2025"
}

@article{Cipriani:2025ikx,
    author = "Cipriani, Andrea and Di Russo, Giorgio and Fucito, Francesco and Morales, Jos{\'e} Francisco and Poghosyan, Hasmik and Poghossian, Rubik",
    title = "{Resumming post-Minkowskian and post-Newtonian gravitational waveform expansions}",
    eprint = "2501.19257",
    archivePrefix = "arXiv",
    primaryClass = "gr-qc",
    doi = "10.21468/SciPostPhys.19.2.057",
    journal = "SciPost Phys.",
    volume = "19",
    number = "2",
    pages = "057",
    year = "2025"
}

@article{Simone:1995qu,
    author = "Simone, Liliana E. and Poisson, Eric and Will, Clifford M.",
    title = "{Headon collision of compact objects in general relativity: Comparison of postNewtonian and perturbation approaches}",
    eprint = "gr-qc/9506080",
    archivePrefix = "arXiv",
    doi = "10.1103/PhysRevD.52.4481",
    journal = "Phys. Rev. D",
    volume = "52",
    pages = "4481--4496",
    year = "1995"
}

@article{Detweiler:1979xr,
    author = "Detweiler, Steven L. and Szedenits, E.",
    title = "{BLACK HOLES AND GRAVITATIONAL WAVES. II. TRAJECTORIES PLUNGING INTO A NONROTATING HOLE}",
    doi = "10.1086/157182",
    journal = "Astrophys. J.",
    volume = "231",
    pages = "211--218",
    year = "1979"
}

@article{Sasaki:1981sx,
    author = "Sasaki, Misao and Nakamura, Takashi",
    title = "{Gravitational Radiation From a Kerr Black Hole. 1. Formulation and a Method for Numerical Analysis}",
    reportNumber = "RIFP-461",
    doi = "10.1143/PTP.67.1788",
    journal = "Prog. Theor. Phys.",
    volume = "67",
    pages = "1788",
    year = "1982"
}

@article{Anninos:1993zj,
    author = "Anninos, Peter and Hobill, David and Seidel, Edward and Smarr, Larry and Suen, Wai-Mo",
    title = "{The Collision of two black holes}",
    eprint = "gr-qc/9309016",
    archivePrefix = "arXiv",
    reportNumber = "PRINT-93-0723 (NCSA,URBANA)",
    doi = "10.1103/PhysRevLett.71.2851",
    journal = "Phys. Rev. Lett.",
    volume = "71",
    pages = "2851--2854",
    year = "1993"
}

@article{Bini:2012ji,
    author = "Bini, Donato and Damour, Thibault",
    title = "{Gravitational radiation reaction along general orbits in the effective one-body formalism}",
    eprint = "1210.2834",
    archivePrefix = "arXiv",
    primaryClass = "gr-qc",
    doi = "10.1103/PhysRevD.86.124012",
    journal = "Phys. Rev. D",
    volume = "86",
    pages = "124012",
    year = "2012"
}

@article{Bini:2017wfr,
    author = "Bini, Donato and Damour, Thibault",
    title = "{Gravitational scattering of two black holes at the fourth post-Newtonian approximation}",
    eprint = "1706.06877",
    archivePrefix = "arXiv",
    primaryClass = "gr-qc",
    doi = "10.1103/PhysRevD.96.064021",
    journal = "Phys. Rev. D",
    volume = "96",
    number = "6",
    pages = "064021",
    year = "2017"
}

@article{Bini:2021gat,
    author = "Bini, Donato and Damour, Thibault and Geralico, Andrea",
    title = "{Radiative contributions to gravitational scattering}",
    eprint = "2107.08896",
    archivePrefix = "arXiv",
    primaryClass = "gr-qc",
    doi = "10.1103/PhysRevD.104.084031",
    journal = "Phys. Rev. D",
    volume = "104",
    number = "8",
    pages = "084031",
    year = "2021"
}

@article{anc,
    author = "Bini, D and Di Russo, G",
    title = "{Associated ancillary file/supplemental material}",
    year = "2026"
}

\end{document}